\documentclass[fleqn,twoside,onecolumn,nofootinbib,showkeys,11pt]{revtex4} %
\usepackage{verbatim}

\usepackage{cmap} %
\usepackage[cp1251]{inputenc}
\usepackage[english]{babel}
\usepackage[T2A]{fontenc}
\usepackage{amsmath}
\usepackage{amstext}
\usepackage{amssymb}
\usepackage{bm}
\usepackage[pdftex]{color,graphicx}%
\usepackage[pdftex,plainpages=false]{hyperref}%
\usepackage{cleveref}%

\begin{document}
\title[Method for identifying crystalline phases in X-ray diffraction data from multiphase samples]
{METHOD FOR IDENTIFYING CRYSTALLINE PHASES IN X-RAY DIFFRACTION DATA FROM MULTIPHASE SAMPLES}%

\author{A.\,D.\,Skorbun}
\affiliation{Institute for Safety Problems of Nuclear
Power Plants,   Nat. Acad. of Sci. of Ukraine}
\address{36a, Kirova str., Chornobyl 07270, Ukraine}
\email{i.zhyganiuk@ispnpp.kiev.ua}
\author{S.\,V.\,Gabielkov}
\affiliation{Institute for Safety Problems of Nuclear
Power Plants,   Nat. Acad. of Sci. of Ukraine}
\address{36a, Kirova str., Chornobyl 07270, Ukraine}
\email{i.zhyganiuk@ispnpp.kiev.ua}
\author{I.\,V.\,Zhyganiuk}
\affiliation{Institute for Safety Problems of Nuclear
Power Plants,   Nat. Acad. of Sci. of Ukraine}
\address{36a, Kirova str., Chornobyl 07270, Ukraine}
\email{i.zhyganiuk@ispnpp.kiev.ua}

\begin{flushleft}
\end{flushleft}

\begin{abstract}
A new method for identifying crystalline phases in X-ray diffraction data has been proposed, which is especially useful for the study of multiphase materials (more than eight --- ten phases) with a relatively low content (less than 1 -- 3 wt\%). The method is based on a statistical analysis of data and provides an unambiguous non-quantitative criterion for the presence of one or another phase in the material. It has been shown that the method works reliably in cases where a significant number of reflexes (more than several dozen) on the diffraction pattern are comparable with intensity-to-noise ratio.  
\end{abstract}


\maketitle
\thispagestyle{empty}

\section{Introduction}

The practice of qualitative X-ray phase analysis (XPA) shows that the application of this method to the analysis of multiphase (more than 8--10 phases) samples, especially with a low (less than 1--3~\%) phase content, is associated with great difficulties~\citep{Skorbun}. This problem is a consequence of the fact that tens or hundreds of reflections, which are formed by a significant number of crystalline phases with a low concentration, are located on the diffraction patterns with high density. The problem becomes more complicated if the sample under study contains phases (for example, aluminosilicates), which produce dozens of reflections of medium and, predominantly, low intensity in their diffraction patterns. Comparison of the experimental data of X-ray diffraction with the data of tables from databases (COD, ASTM, and others) does not make it possible to enumerate hundreds of proposed versions for identifying these phases. Using the code ''Match!''~\citep{Brandenburg} also turns out to be ineffective for identifying phases in such specific samples. Thus, the patterns of such multiphase samples contain a combination of reflections from many phases, in which, in addition, some low-intensity reflections may be absent, and intense reflections may appear due to the superposition of some reflections from different compounds.

In the paper~\citep{Skorbun} it was shown that the method of analysis of noisy signals, developed in~\citep{PanasyukSkorbun} and based on the methods of computational statistics~\citep{Moore}, makes it possible to distinguish weak in intensity reflections (at the noise level) on diffraction patterns of materials with a low content of crystalline phases. 

In this work, the method of searching for individual reflexes in ''noisy'' patterns, proposed in~\citep{Skorbun, PanasyukSkorbun}, based on the use of statistical analysis, is developed for the problem of finding a combination of reflexes in such complicated patterns. In other words, a method has been developed for automated identifying in the experimental diffraction pattern of the material under study, the contributions of individual phases with their small relative content in the material under study.

\section{Correlation method for diffraction pattern analysis (phase search)}

In the practice of X-ray phase analysis, there are empirical criteria for identifying crystalline phases from X-ray diffraction data, which remain valid until now~\citep{Dinnebier, Gilmore} and which, in fact, can be used as an algorithm for developing the corresponding software.
\begin{enumerate}
	\item The positions (an angle 2~$\theta$) of the reflections (from 3 to 6 reflections) in the diffraction pattern should coincide with the positions of the reflections of the assumed crystalline phase taken from the corresponding databases, the presence of which in the sample we are trying to establish.
	\item These~3--6 reflections should be the most intensive among other possible reflections of the same phase in the X-ray diffraction data from the sample under study.
	\item These~3--6 reflections should be located (following their relative intensity) in the same sequence in which they are located for the same phase in the X-ray diffraction database. 
\end{enumerate}

As one can see, all of the above criteria are based on a visual comparison of experimental diffraction patterns with diffraction patterns based on X-ray diffraction data from databases. However, in reality, this approach is not applicable for materials in which more than 4--6 phases and, accordingly, 20--90 reflections are present. Because individual reflexes from different phases may coincide, it is difficult to determine whether a specific reflex of medium intensity is one of 6 reflexes from a phase with the highest content, or, this is the main (i.e., the first) reflex from a phase with a lower content in the investigated sample. The situation becomes more complicated when it is taken into account that both the position of the reflections (angle 2~$\theta$) and their intensities have an experimental error.

It is difficult to identify the phase when, due to experimental error, the intensity of the reflex (which is the average in a series of 6 reflections of one phase) is equal to or greater than the intensity of the first reflex of the other phase. The absence of individual reflexes of low intensity against the background noise can also be considered as a situation when a real diffraction pattern has its structure of reflexes only with a certain probability.

Known programs for processing XRD data~\citep{Brandenburg, Gilmore} give satisfactory results in cases of studying materials with a small number of phases and give practically unacceptable results in cases where the intensities of reflections are at the noise level. Thus, it becomes clear that for the process of phases identifying in materials with a significant amount of them, it is advisable to create and use specialized mathematical software products with algorithms built on the methods of computational statistics, which will allow identifying phases in the above-defined case.

The traditional way of the diffraction pattern processing, described above~\citep{Toby}, is essentially deterministic. This paper considers the possibility of using probabilistic methods. 

Since the angles for the patterns are known exactly, especially for those given in the databases, at first glance it may seem that it is necessary to compare the intensities of the reflexes for the given angles. However, in reality, the positions of the reflections (angles) in the X-ray phase analysis are determined with some error. The values of the relative intensities of the reflexes also have their own uncertainty ($\pm 15-25$\%), which is difficult to take into account with the traditional identification method. That is, here, too, the result is obtained with a certain degree of reliability. And, therefore, it is implicitly assumed that the data is to some extent random.

The discussion about the traditional diffraction pattern processing raises the need to consider the new approach, instead of the three criteria formulated above. New methods would make it possible to estimate the reliability of the conclusion about the phase identification. In a stricter statistical formulation: estimate the probability of a type 1 error --- the adoption of an incorrect hypothesis~\citep{Pollard}. Standard software products of the type~\citep{Brandenburg} applied to multiphase materials give a hard-to-see array of assumed phases, which requires an unacceptable investment of time for their analysis. Therefore, it is logical to formulate the problem of increasing the efficiency of the identification process using statistical methods.

To solve the problem of searching for a previously established combination of reflections among many reflections on an experimental diffraction pattern, it is proposed to search for a correlation between samples of experimental data (on the basis of which the experimental diffraction pattern was built) and X-ray diffraction data taken from various powder-diffraction experiments data. The calculations are based on the use of computational statistics methods to determine correlations~\citep{Skorbun, PanasyukSkorbun, Moore}. First, let's reformulate the problem in terms of statistics.

\begin{figure}[h]
\vskip1mm
\begin{center}
\includegraphics[width=0.80\textwidth]{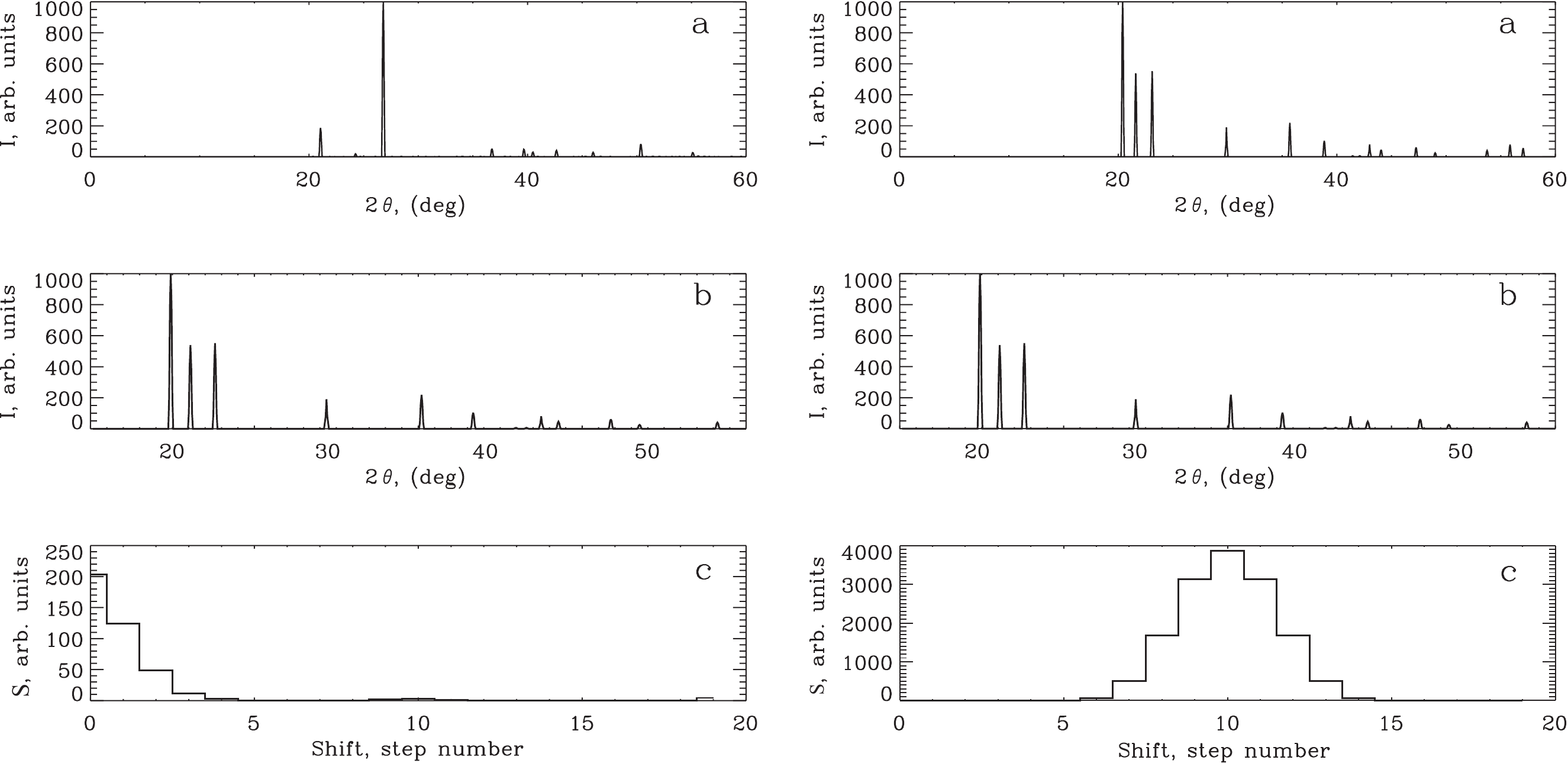}
\end{center}
\vskip-1mm
\caption{Results of search of diffraction pattern from database in a ''model'' diffraction pattern. Left:  search for a tridymite in diffraction pattern of the same tridymite: a – ''model'' tridymite diffraction pattern; b – tridymite diffraction pattern from database; c – resulting histogram of the rate of correlation between them. Right: result of the same search of a tridymite in an $\alpha$-quartz.
}\label{f1}
\end{figure}

\begin{enumerate}
	\item The diffraction pattern as a whole or some part of it (selected range of angles) will be considered as a statistical sample. In this case, the angles are taken without gaps regularly with a measurement step, say, in $0.05^{\circ}$. Each angle value corresponds to a certain intensity value, that is, these pairs of values~$\left\{2 \theta_i, I_i\right\}$  form a regular sample of pairs of values.
	\item From the database~\citep{Cedric}, a table of angles and relative intensities is taken, which correspond to the assumed phase, the presence of which must be determined in the experimental diffraction pattern. In these tabular data, there are only individual angles and the corresponding values of the relative intensities. Therefore, they need to be transformed into a form that has an experimental diffraction pattern. For this, a sample is created with the same dimension as the experimental one in the previous paragraph, that is, with the same values of the angles, but with zero intensities. Further, the zero values of the intensities for the tabular (from the database~\citep{Cedric}) angles are replaced with the corresponding tabular values of the relative intensities from the same database. Thus, a sample of intensities for tabular angles is created, with the same dimension as the experimental one.
	\item The databases provide a specific value of the angle for the corresponding reflex. But this is only the value of the angle at the maximum of the reflex, which can, moreover, have different values according to the data of different authors (as an example, see the analysis of the situation in~\citep{Skorbun}). In the practice of X-ray phase analysis, the difference between the experimental values of the angles and the corresponding values of the angles from the databases by an amount of $0.004 - 0.02^{\circ}$ is considered acceptable. In the method described below, the values of the intensities of the experimental and tabular (from the XRD databases) samples for the corresponding angles are multiplied. If the angles differ by the chosen step of $0.05^{\circ}$, the result of multiplication will be equal to zero, since in the experimental or the tabular samples they will have zero values of the intensities.
	\item To take into account the widths of the reflexes, in the sample created from the database, it is necessary to convert single tabular values for certain angles into reflections of a certain width. The question of the shape of the reflection constructed on the basis of experimental X-ray diffraction data~\citep{Rozhenko} requires separate consideration. The shape of the reflections, however, is not fundamental for our purposes to create a method able to look for set of reflections which corresponds to the desired crystalline phase from the database, among the reflections of the experimental diffraction pattern. Therefore, one of the simplest ways to 'broaden' the tabular reflex, built on the basis of data from the XRD databases, is to take the values for the shape of the reflexes from the experimental pattern, or simply arbitrarily assign half the intensity values for the intensity values to the left and right of the tabular angle value.
	\item After creating such samples from experimental data and data from XRD databases for the desired phase, it is necessary to consider the task of their statistical comparison, namely, to calculate the degree of correlation between them. We note right away that the calculation of the usual correlation coefficient (Pearson or others) does not give an answer to the questions posed. That is why a method is proposed for calculating the degree of correlation between the created samples, which is based on the methods of computational statistics~\citep{Moore, Rozhenko}). These methods, in turn, are based on the use of Monte Carlo methods. In-depth substantiation of the need to apply computational statistics approaches for our problem of analysing patterns, algorithms and formulas can be found in~\citep{Skorbun, PanasyukSkorbun, Moore, Rozhenko}). This analysis method is based on calculation of correlation. So, if you take two identical samples of diffraction data, these data will be fully correlated. A feature of the search for correlation in the proposed method is that the diffraction data of the phase, the presence of which is established in a complex experimental pattern, are not a sample that must be scanned over different parts of the experimental diffraction pattern - the total range of angles is the same for both patterns. The question is how much the intensities for a given set of angles are correlated. At first glance, it seems that after creating two samples for comparison, as described above, you can simply calculate the correlation coefficient between them.
	\item However, direct calculation of the degree of correlation between two data samples displayed on two diffraction patterns turned out is ineffective in the sense that the obtained correlation coefficient has nothing to compare with. We admit a priori that both for various reasons, listed above, and from simple considerations that in noisy diffraction patterns, the ratios of the intensities of reflexes are distorted by noise comparable in intensity with the tabular data for reflexes. For this reason, we cannot hope to get a 100\% correlation. And there is no way to calibrate the calculation results. Therefore, the following technique was proposed. The experimental dataset and the dataset from the X-ray diffraction databases are considered as two samples with identical angles, but different intensities for these angles. Let us shift the data set of the intensities of the diffraction pattern of the sought phase from the database by a certain number of positions to the left along the abscissa, on which the values of the angles are plotted (in the examples below, ten steps were chosen, which corresponds to a shift of $10 \times 0.05 = 0.5$ degrees) (see Figure~\ref{f1}). Now, using the above-mentioned methods of computational statistics, we calculate the correlation between the experimental set of diffraction data and database data corresponding to the desired phase, each time one step shifting towards the ''correct'' correspondence of angles, totally on twenty steps. In the case when we calculate the correlation of two identical datasets, we get the maximum correlation at the $10^{\rm th}$ step, which will begin to decrease further with the growing of differences between angles. Note that at some shift of angles of the diffraction patterns, a noticeable correlation is also possible, but we know (since this is included in the calculation method) that the true result is obtained only at the $10^{\rm th}$ step.
\end{enumerate}

A feature of computational statistics is that the evaluated correlation degree is not a correlation coefficient in parts of unity, but the probability of correct acceptance of the null hypothesis about the identity of the X-ray diffraction data set from the databases and the experimental data set. At the same time, due to the specifics of the method, the result of calculations is obtained in the form of some relative units. This still does not make it possible to compare the relative content of individual crystalline phases with each other in a sample, but makes it possible to reliably establish the presence of a given phase in the same sample by the degree of correlation between the data sets of the experimental and tabular patterns. A non-zero correlation between the experimental dataset and the selected phase dataset from the database indicates this fact. If someone wants to automate the process of deciding about the identification of a phase, he will also have to establish a criterion for this. No non-empirical methods for comparing such datasets have been developed to date. In our case, this means that the experimenter himself decides whether the peak at a step 10 appears on the histogram reliably.

\section{Demonstration calculations}

First, we will demonstrate the above mentioned by the example of calculating the degree of correlation between samples from the X-ray diffraction database. Note that the intensity values for these data are normalized so that the maximal intensity reflex is set to 1000. As examples, silicon oxides tridymite and cristobalite were arbitrarily chosen. To begin with, to demonstrate the operation of the recognition algorithm, let us compare the same sample of X-ray diffraction data for tridymite from the COD database (Figure~\ref{f1}~a). We will consider one sample, which is displayed as the diffraction pattern (Figure~\ref{f1}~a), obtained experimentally, while it, shown as a diffraction pattern in Figure~\ref{f1}~b will be a ''test''.

Figure~\ref{f1}~c shows the result of the analysis, which is constructed as the result of calculating the degree of correlation between two samples using the scanning algorithm as described above. The degree of correlation is calculated in relative units because it depends on the absolute values of the intensities in the samples, which are compared with each other~\citep{Moore, Gilmore}. And the very fact of correlation is evidenced by the fact that at a step 10 there is a peak whose height exceeds all others and is given in these relative units. As expected, the maximum correlation is observed when the offset between the angles is zero, that is, at the $10^{\rm th}$ scan step.

The same figure shows a histogram of the degree of coincidence between the $\alpha$-quartz diffraction pattern and the cristobalite diffraction pattern. This situation, when the diffraction patterns of the compared compounds are different, should demonstrate that in such an obvious case the degree of correlation is zero --- there is no peak at a step 10. We emphasize that it is impossible to compare the heights of the peaks for different phases at this stage of calculations:  as it was said, they depend on the values of the intensities of reflexes in the initial patterns.

Therefore, in Figure~\ref{f1} it is clearly seen that between the ''experimental'' (a) and tabular (b) patterns (in this case, identical), there is a reliable correlation, which is manifested by the presence of a peak in the histogram (c) at $10^{\rm th}$ step. And it is obviously seen, that there is no correlation between different diffraction patterns ($\alpha$-quartz and tridymite).

\section{Analysis of complex model patterns}

\begin{figure}[h]
\vskip1mm
\begin{center}
\includegraphics[width=0.40\textwidth]{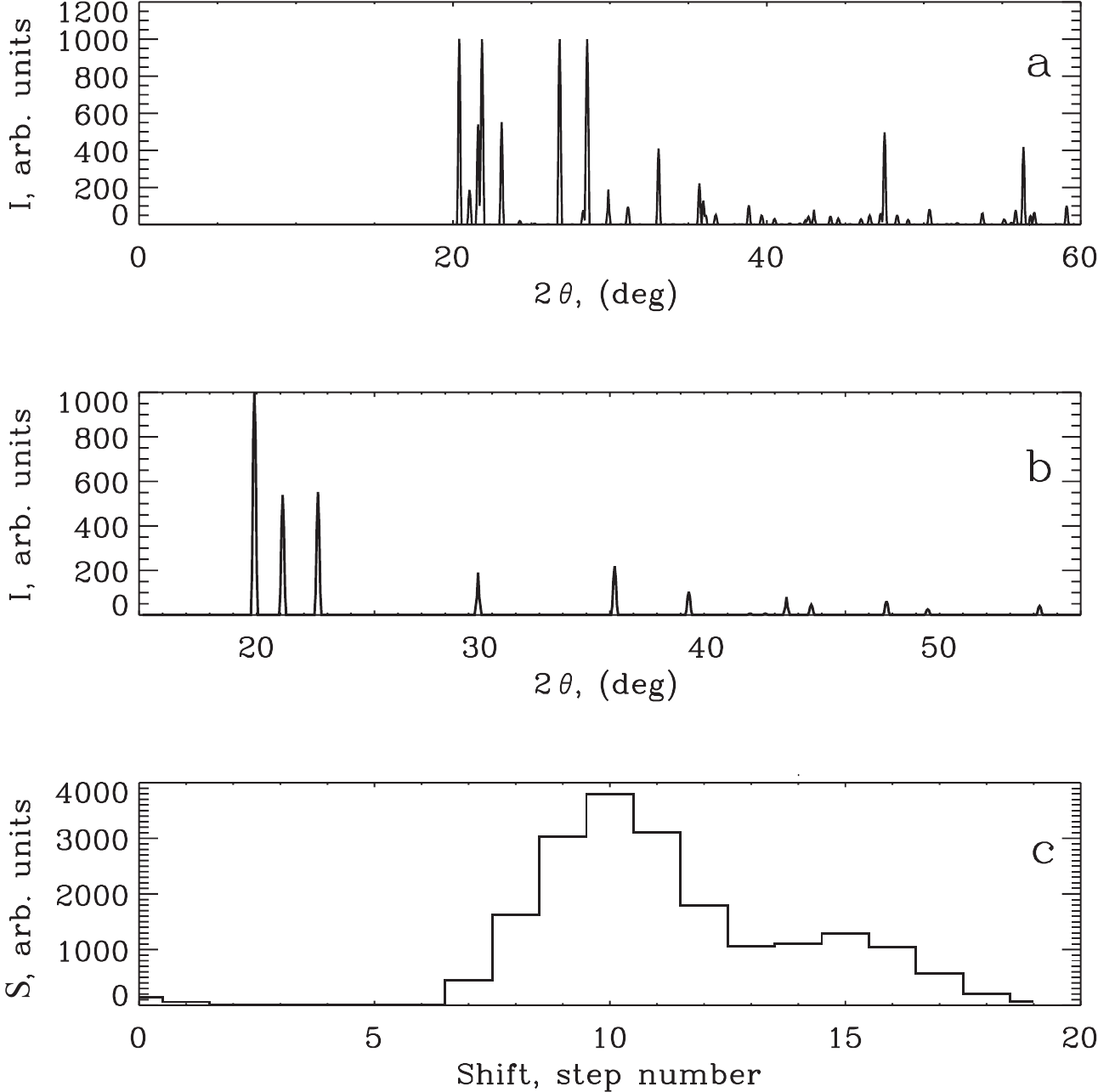}
\end{center}
\vskip-1mm
\caption{Results of search for tridymite diffraction pattern from database in a ''model'' diffraction pattern. a -- ''model'' diffraction pattern; b -- tridymite diffraction pattern from database; c -- resulting histogram of the rate of correlation between them.}\label{f2}
\end{figure}

\begin{figure}[h]
\vskip1mm
\begin{center}
\includegraphics[width=0.40\textwidth]{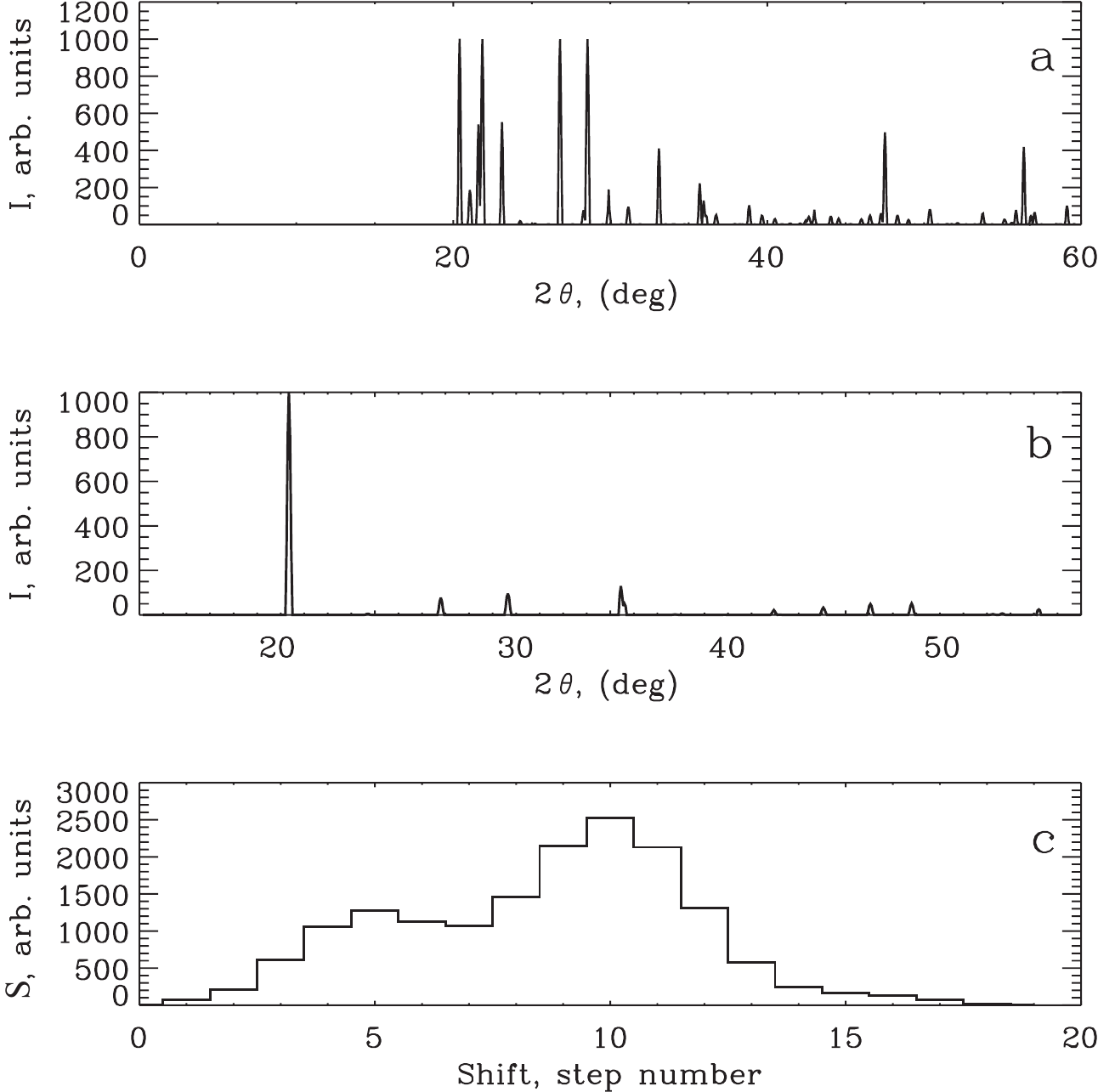}
\end{center}
\vskip-1mm
\caption{Results of search for cristobalite diffraction pattern from database in a ''model'' diffraction pattern. a -- ''model'' diffraction pattern; b -- cristobalite diffraction pattern from database; c -- resulting histogram of the rate of correlation between them.}\label{f3}
\end{figure}

\begin{figure}[h]
\vskip1mm
\begin{center}
\includegraphics[width=0.40\textwidth]{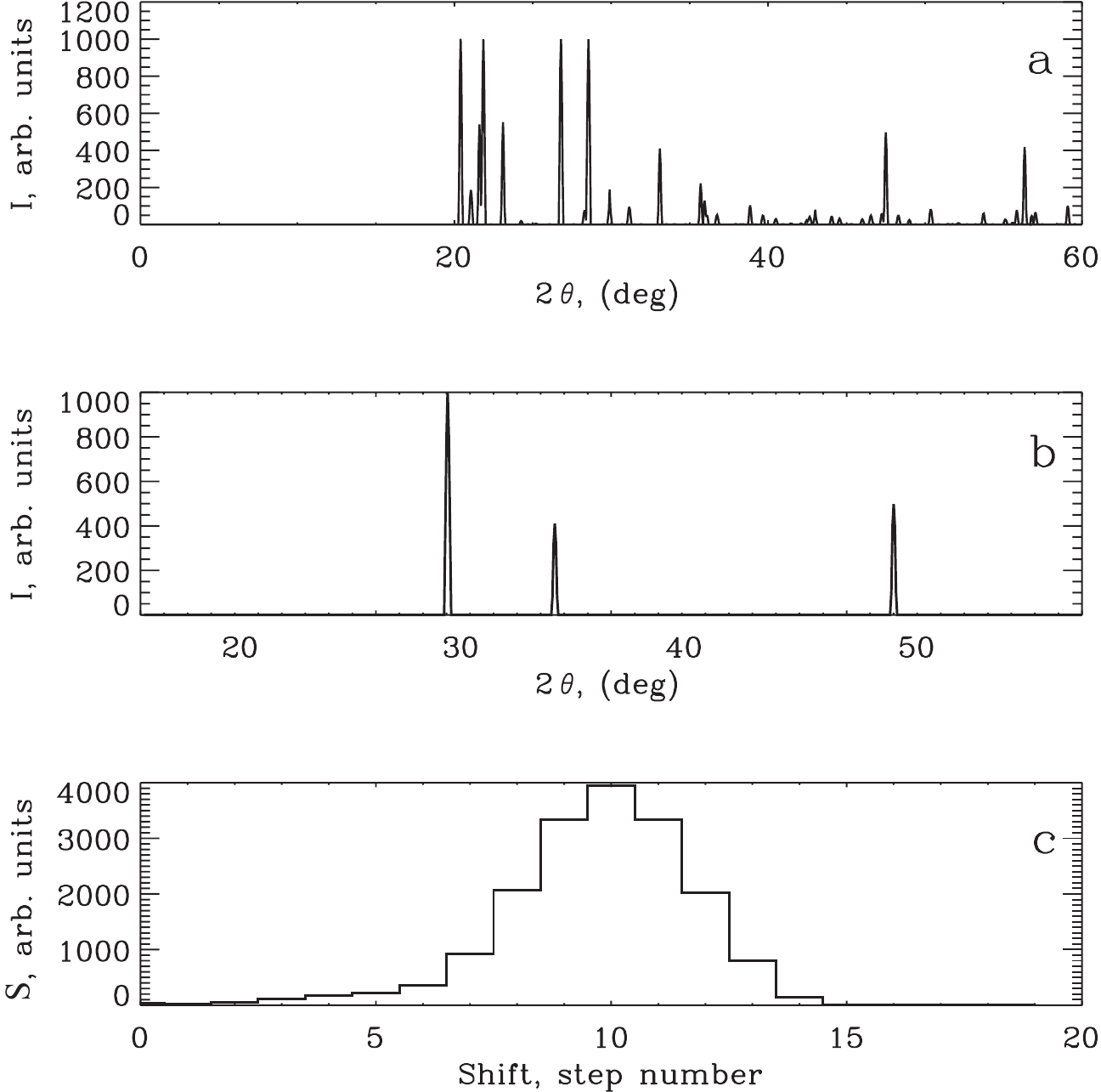}
\end{center}
\vskip-1mm
\caption{Results of search for dioxide uranium diffraction pattern from database in a  ''model'' diffraction pattern. a -- ''model'' diffraction pattern; b -- dioxide uranium diffraction pattern from database; c -- resulting histogram of the rate of correlation between them.}\label{f4}
\end{figure}

\begin{figure}[h]
\vskip1mm
\begin{center}
\includegraphics[width=0.40\textwidth]{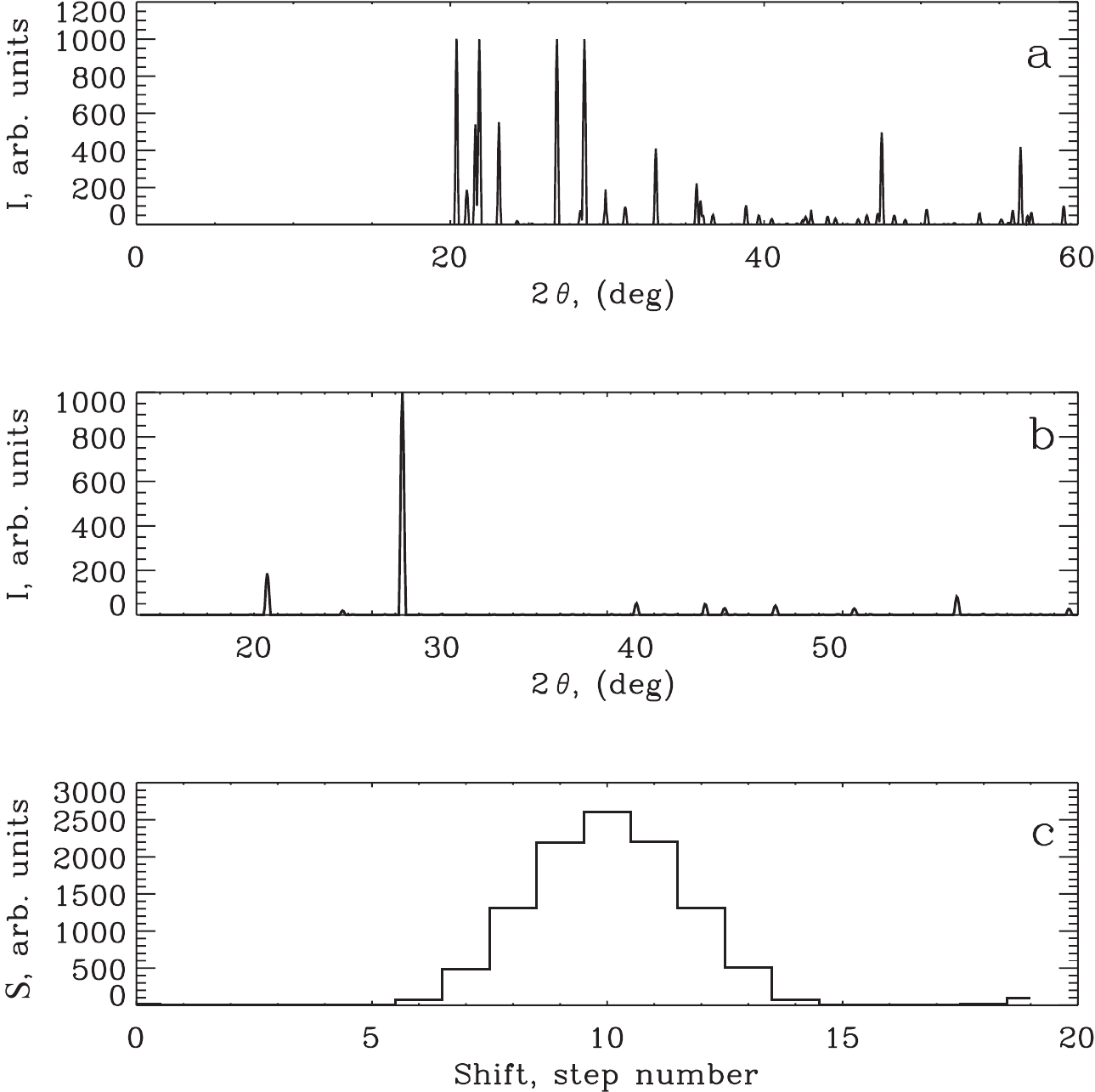}
\end{center}
\vskip-1mm
\caption{Results of search for $\alpha$-quartz diffraction pattern from database in a ''model'' diffraction pattern. a -- ''model'' diffraction pattern; b -- $\alpha$-quartz diffraction pattern from database; c -- resulting histogram of the rate of correlation between them.}\label{f5}
\end{figure}

Now let's check the proposed method with a more complex example. This example is also a model example because if we are testing a method, we must know the correct answer before starting work. Therefore, we take tabular diffraction patterns of tridymite, cristobalite, uranium dioxide, and diffraction pattern of $\alpha$-quartz. Let's combine these four diffraction patterns into one by adding arrays element-wise and apply our method to find the four phases in the resulting diffraction pattern. The result is shown in Figures~\ref{f2},~\ref{f3},~\ref{f4},~\ref{f5}. For all search options, there is a peak at a step 10 of the scan, that is, all phases, that were included in the total diffraction pattern, were found.

\section{Model diffraction pattern --- the sum of experimental data and data from the powder diffraction database}

Because our work is methodical in its nature, then the results of the suggested method must be shown on the examples, where results are known in advance. But such data are practically absent: the databases have information only about separate compounds, but not about a mixture of phases.

It turned out to be not so easy, to find a test example of a complex experimental diffraction pattern to demonstrate the real capabilities of the proposed method. Thus, among the lava-like radioactive materials of the destroyed $4^{\rm th}$~power unit of the Chornobyl NPP the crystal phases are located in the depleted glass phase. As a result, the reflexes on the diffraction patterns appear to be a little shifted from their tabular values due to the deformation of the lattice due to the appearance of mechanical stresses in the glass phase~\citep{Gabielkov}. So the diffraction patterns of such materials are different from the table data and cannot be serviced as tests due to they themselves require non-standard identification. Nevertheless, there is a desire to demonstrate how the method works on real data.

Thus, the only way to receive a complicated diffraction pattern for testing is to model them, similarly to made above by summing numbers of diffraction patterns, but more complicated ones. The experimental diffraction pattern of the so-called ''brown'' lava-like materials of the $4^{\rm th}$~Unit of the ChNPP was taken as the basic. It has an appearance of a noise band with some weak reflexes. But it is known, however, there are many phases in this material~\citep{Zhyganiuk}.

\begin{figure}[h]
\vskip1mm
\begin{center}
\includegraphics[width=0.80\textwidth]{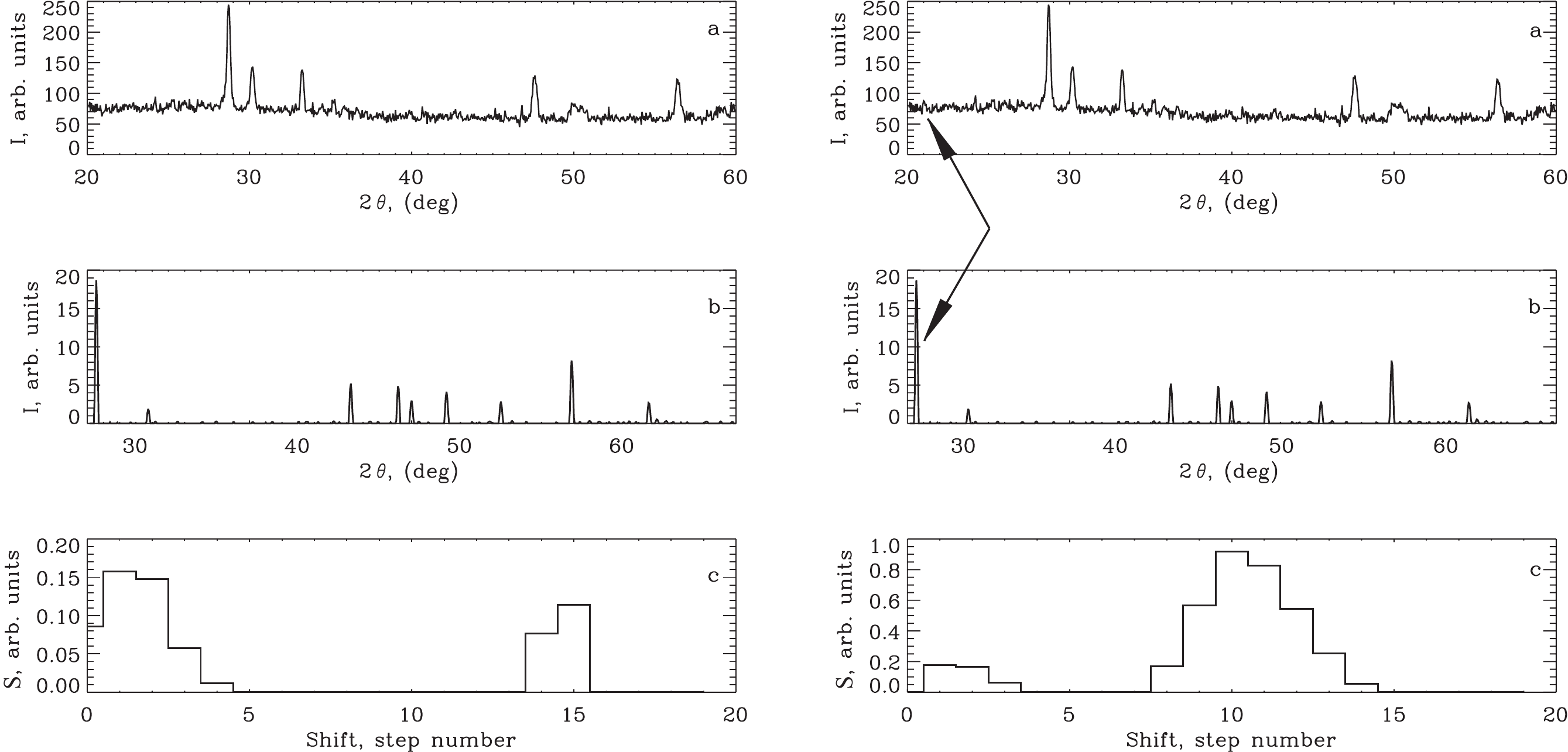}
\end{center}
\vskip-1mm
\caption{Results of the same search of an $\alpha$-quartz in sum of an $\alpha$-quartz and brown ceramics. Left:  search for an $\alpha$-quartz in diffraction pattern of brown ceramics: a -- brown ceramics diffraction pattern; b -- an $\alpha$-quartz diffraction pattern from database; c -- resulting histogram of the rate of correlation between them. Right: result of the same search of an $\alpha$-quartz in sum of an $\alpha$-quartz and brown ceramics.
}\label{f6}
\end{figure}

As an addition to this diffraction pattern, also the experimental diffraction pattern from the stsndard of an $\alpha$-quartz was chosen, from which the most intense reflex ($2 \theta = 25^{\circ}$) was deleted. In such a model construction the intensity of maximal line (the second line $2 \theta = 21.05^{\circ}$ of $\alpha$-quartz, see Figure~\ref{f6}~b) was made about 0.1 from the maximal line of the based diffraction pattern to check the sensitivity of our method to weak reflexes. Then these two diffraction patterns were combined in one (Figure~\ref{f6}~a). In such a combined diffraction pattern the line from $2 \theta = 21.05^{\circ}$ is hardly seen (see arrow), and weaker lines of an $\alpha$-quartz are not always visible above the noise, but we will remember, that they are presented there.

The result of our searching for an alpha-quartz in such a diffraction pattern is shown in Figure~\ref{f6}~c. The histogram with the clear peak on a $10^{\rm th}$ step demonstrates, that the developed method satisfactory works in such a noisy situation. 

\section{An example with a real diffraction pattern}

Above, the efficiency of the method was demonstrated on model data, which can be considered as an example of relatively complex patterns. However, of much greater interest is the question of the capabilities of the method for analyzing data with a significantly larger set of phases, the total effect of reflexes from which is already similar to noise. 

\begin{figure}[h]
\vskip1mm
\begin{center}
\includegraphics[width=0.40\textwidth]{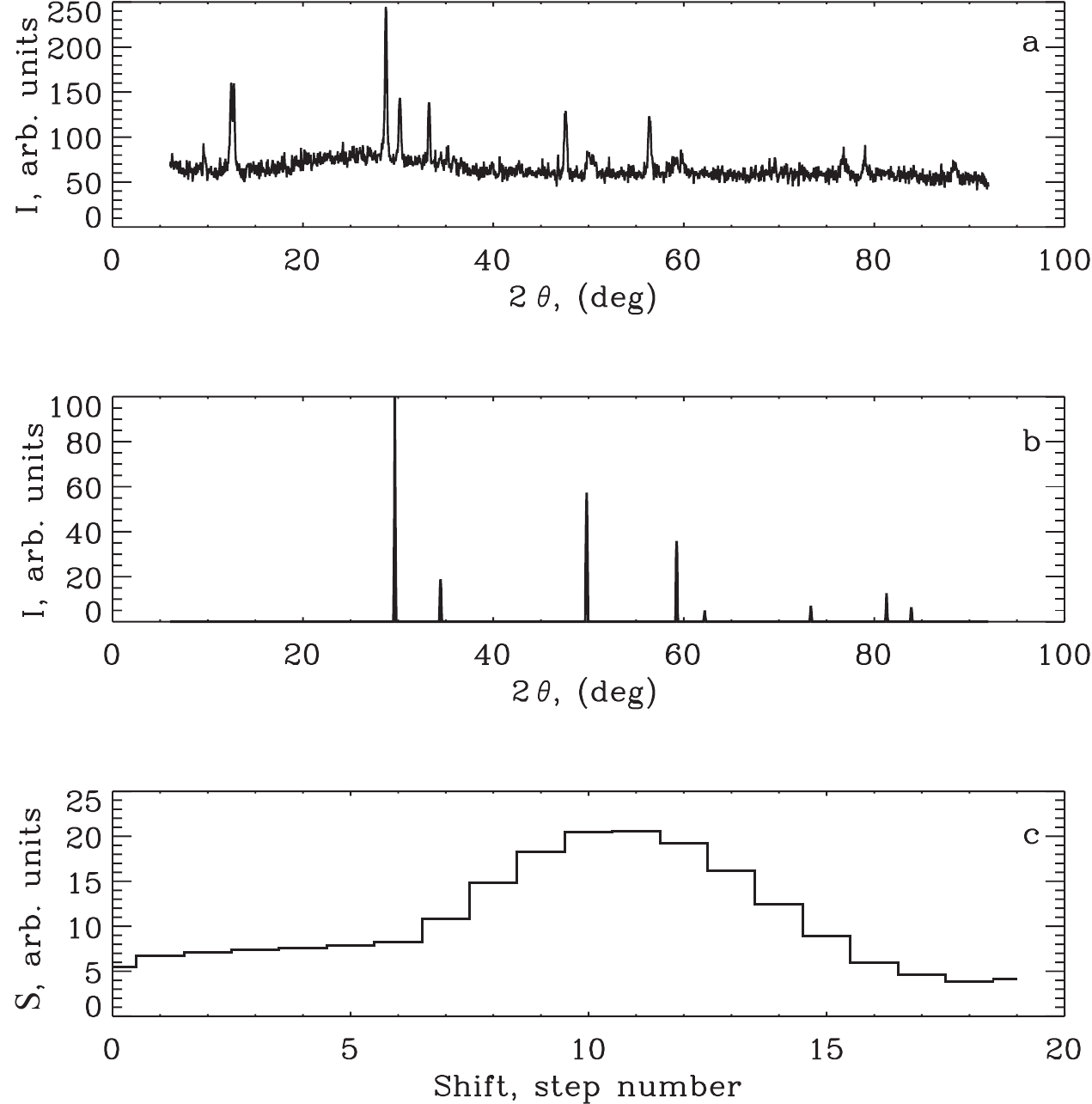}
\end{center}
\vskip-1mm
\caption{Results of search for a  cubic zirconium oxide in diffraction pattern of brown ceramics. a -- brown ceramics diffraction pattern; b -- cubic zirconium oxide diffraction pattern from database; c -- resulting histogram of the rate of correlation between them.}\label{f7}
\end{figure}

As such an experimental example, the diffraction pattern of so-called brown lava-like fuel containing materials of the Unit $4^{\rm th}$ of the Chornobyl NPP from the Figure~\ref{f6} was taken also, but without adding an $\alpha$-quartz. Externally it looks like a noise band with several weak reflections, that can be attributed to uranium oxide. It is well known, however, that these lavas have a complex phase composition~\citep{Gabielkov}. From various data~\citep{Gabielkov, Borovoi}, it was established that these lavas should contain zirconium oxide, the presence of which we will try to verify by the developed method. The result of calculating the degree of correlation between diffraction patterns of brown ceramics and cubic ${\rm Zr O}_2$ is shown in Figure~\ref{f7}. In this Figure, it can be seen that the presence of cubic ${\rm Zr O}_2$ is reliably established by the presence of a peak at a step 10 of the search for correlations. In this case, only the first from all the ${\rm Zr O}_2$ reflections is visually visible against the background of the diffraction pattern. Note that the value of the correlation at the maximum of the histogram in Figure~\ref{f7} (30) is much smaller than that in Figures~\ref{f3}--\ref{f5} for model patterns. This is fully consistent with the expectations since the expected reflections from ${\rm Zr O}_2$ are not visible in the experimental diffraction pattern against the background noise. This means, that its concentration is very small.

\section{Conclusions}

The method for phases identification in diffraction patterns, which is based on using a statistical approach has been developed. The method uses the calculation of a correlation between real and tabular diffraction patterns.

The method, proposed in the presented work for the identification of crystalline phases in multiphase samples allows:
\begin{description}
	\item[a)] reliably to identify the presence of certain crystalline phases in multiphase samples;
	\item[b)] the method has increased sensitivity and makes it possible to identify phases of relatively low content in noisy patterns, namely: reliably to identify phases, whose reflex intensities are commensurate with the background;
	\item[c)] at this stage, the developed method for identifying phases in noisy patterns only answers the question about the presence of this phase in a complex pattern. It gives an unambiguous, although still the non-quantitative criterion for the presence of one or another phase: the presence of a peak in the histogram for the correlation coefficient, which is in the correct position (in our case, at the $10^{\rm th}$ step of the angle shift). At the same time, we emphasize that the presented results indisputably indicate that the method works reliably where other methods do not work at all, namely, when the reflexes are practically invisible against the background of noise. That is, as the final result, the use of the proposed method for determining correlations increases the reliability of phase identification.
\end{description}

The question of quantifying the content of the identified phase by the proposed method requires further research.

The work was sponsored in the framework of the budget theme of the National Academy of Sciences of Ukraine
(No. 0221U100249 ).

\end{document}